# Period measurement by accelerating observers

Bernhard Rothenstein[1)] and Stefan Popescu [2)]
1) Politehnica University of Timisoara, Physics Department, Timisoara, Romania
2) Siemens AG, Erlangen, Germany

**Abstract.** *We consider the problem of the period measurement in the case of the following scenarios: stationary source of successive light signals and accelerating receiver, stationary receiver and accelerating source of successive light signals and stationary machine gun that fires successive bullets received by an accelerating receiver. The accelerated motion is the hyperbolic one.*

## 1. Introduction

A periodical phenomenon is characterized by its period that is the time interval between its successive occurrences. By definition a clock generates a periodical phenomenon characterized by a rigorously constant period. A clock at rest at some point of the plane defined by the axes of the inertial reference frame K(XOY) generates the events $E_1(x,y,t)$ and $E_2(x,y,t+\Delta t)$ by two successive ticks. Here (*x,y*) are the space coordinates of the point where the clock is located and $\Delta t$ is the time interval between two successive ticks called proper period of the clock because it is measured as a difference between the readings of the same clock located at the point where the two events take place. Clocks play a fundamental part in Einstein's special relativity. It is considered that at each point of the plane defined by the axes of an inertial reference frame we find a clock, all these clocks displaying the same running time as a result of their synchronization in accordance with the synchronization procedure proposed by Einstein.[1] We are interested only in the clocks located at the different points of the OX axis, an important role being played by a clock $C_0(0,0)$ located at the origin O and by a clock $C'_0$ that moves with speed *V* in the positive direction of the OX axis reading $t'=0$ when it passes in front of clock $C_0$. We present in Figure 1 the situation when clock $C'_0$ reading *t'* is located in front of a clock $C(x=Vt, y=0)$ reading *t*. The readings of the two clocks are related by

$$t = \frac{t'}{\sqrt{1-V^2/c^2}} \qquad (1)$$

as a consequence of the principle of relativity[2]**.**

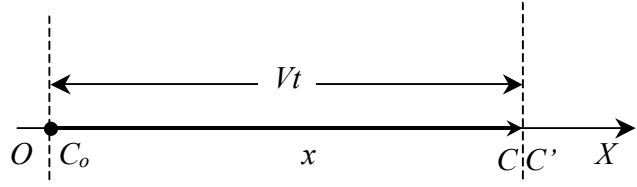

**Figure 1**. The relative position of the synchronized stationary clocks $C_0(0,0)$, $C(x=Vt, y=0)$ and of the moving clock C' at a given time as detected from the K frame

The constancy of $V$ enables us to present (1) as
$$\Delta t = \frac{\Delta t'}{\sqrt{1-V^2/c^2}} \qquad (2)$$
Here $\Delta t' = t' - 0$ represents a **proper time interval** being measured as a difference between the readings of the same clock $C'_0$. Further $\Delta t = t - 0$ represents a **non-proper time interval** being measured as a difference between the readings of two different clocks $C(x=Vt, y=0)$ and $C_0(0,0)$ when clock $C'_0$ passes in front of them. Equation (2) accounts for the time dilation effect. The constancy of $V$ ensures the time independence of (2).

Consider a source of light $S(0,0)$ located at the origin O of the K frame. When the clock $C_0(0,0)$ reads $t_e$ the source emits a light signal in the positive direction of the OX axis. The light signal arrives in front of a clock $C(x,0)$ when it just reads $t_r$. It is obvious that
$$t_r = t_e + x/c. \qquad (3)$$
By differentiating both sides of (3) we obtain
$$dt_r = dt_e + dx/c. \qquad (4)$$
Let $C'$ be a clock moving with speed $V$ located in front of clock $C(x,0)$ when the light signal arrives at its location. We consider that
$$dx/dt_r = V \qquad (5)$$
represents the speed of clock $C'$.

In the equations (3,4) above $dt_e$ represents the period at which source $S$ emits successive light signals and $dt_r$ the period at which an observer commoving with clock $C'$ receives the successive light signals, both $dt_e$ and $dt_r$ being measured in the rest frame of the source $S$. By definition $dt_e$ represents a proper time interval with $dt_r$ being a non-proper time interval. The same time interval $dt'_r$ measured as a difference between the readings of clock $C'$ is a proper time interval and it is related to $dt_r$ by (2)

$$dt_r = \frac{dt'_r}{\sqrt{1-V^2/c^2}} \tag{6}$$

With the new notations (4) leads to
$$dt'_r = dt_e \sqrt{(1+V/c)/(1-V/c)}. \tag{7}$$

Equation (7) accounts for the Doppler Effect, establishing a relationship between the proper time interval $dt_e$ at which the stationary source of light emits the successive light signals and the proper time interval $dt'_r$ at which a receding observer receives them. If the observer approaches the source then we will have
$$dt'_r = dt_e \sqrt{(1-V/c)/(1+V/c)}. \tag{8}$$

As long as the velocities involved are constant and the motions take place along the line that joins source and observer (longitudinal Doppler Effect) (7) and (8) hold for finite proper time intervals $\Delta t'_r$ and $\Delta t_e$ respectively.

The problem is to find out what happens if the involved motions are not uniform[3,4,5,6,7]. The purpose of our paper is to extend the results known so far and to underline the difference between the different approaches. We use the subscript *e* for emission and *r* for reception with primed symbols to show that accelerating clocks measure the corresponding quantities and un-primed symbols in order to show that stationary clocks measure these quantities.

Equation (7) tells us that when the observer is receding the source increasing his velocity the result is an increase of $dt'_r$, whereas a reduction of the velocity is accompanied by a reduction of the considered time interval. Equation (8) tells us that when the observer approaches the source increasing his velocity the result is a decrease of $dt'_r$, whereas a reduction of the velocity is accompanied by an increase of the same time interval.

We consider the hyperbolic motion previously known from [8]. This is a particular motion that begins at $x = \infty$, $t = -\infty$ with $V = c$ and proceeds through a first deceleration phase when the object approaches the origin $x = 0$ with proper acceleration $-g$ until the it reaches the still stand ($V = 0$) at $t = 0$ and $x = x_0$. Thereafter for $t > 0$ the moving object enters a second acceleration phase receding the origin with the proper acceleration $+g$ reaching $V = c$ as $t \to +\infty$ in accordance with the requirements of special relativity. The deceleration and acceleration phases are separated by $t = t' = 0$ at $x_0 = c^2/g$. The distance travelled by an accelerating clock may be expressed as a function of constant proper acceleration $g$ and either the reading $t$ of a stationary clock or the reading $t'$ of the accelerating clock located in front of the stationary one as follows

$$x = \frac{c^2}{g}\left(\sqrt{1+g^2t^2/c^2}\right) \quad resp. \quad x = \frac{c^2}{g}\cosh(gt'/c) \qquad (9)$$

The instantaneous velocity of the accelerating clock is

$$V = \frac{gt}{\sqrt{1+\left(\frac{gt}{c}\right)^2}} \quad resp. \quad V = c\tanh(gt'/c) \qquad (10)$$

The clock readings are related by

$$t = \frac{c}{g}\sinh(gt'/c). \qquad (11)$$

Unlike previously known approaches [8,3,5] we will use the novel notation $T = \frac{c}{g}$ in order to simplify the equations and to better underline the physical meaning of various terms. Here $T$ is a "magic" time interval required to accelerate an object with constant acceleration $g$ until it reaches the speed $c = gT$ in the non-relativistic way. With this new notation we have $x_0 = cT = gT^2$ and the previously known equations (9) and (10) changes to

$$x = \frac{x_0 + gt^2}{\sqrt{1+\left(\frac{t}{T}\right)^2}} \quad resp. \quad x = cT\cosh\left(\frac{t'}{T}\right) \qquad (9a)$$

$$V = \frac{gt}{\sqrt{1+\left(\frac{t}{T}\right)^2}} \quad resp. \quad V = c\tanh\left(\frac{t'}{T}\right) \qquad (10a)$$

When $|t| \ll T$ or $|t'| \ll T$ these expressions reduce to the well known equations of the non-relativist accelerated motion. The relationship between the clock readings changes as follows:

$$t = T\sinh\left(\frac{t'}{T}\right) \quad or \quad \frac{t}{T} = \sinh\left(\frac{t'}{T}\right). \qquad (11a)$$

Further we identify the particular time window $|t'| \ll T$ when $t \cong t'$ and thus $|t| \ll T$. Within this temporal window the relativist effects may be neglected and the hyperbolic motion may be approximated by the non-relativistic accelerated motion.

Subsequently we will consider the following scenarios:
   a. Stationary source of light and accelerating receiver.
   b. Stationary observer and accelerating source of light.
   c. Stationary source of acoustic waves and accelerating receiver.

**2. Stationary source of light and accelerating receiver**

The source of monochromatic light, located at the origin O of the inertial reference frame K emits successive very short light signals (wave crests) at a constant proper period $T_e$.

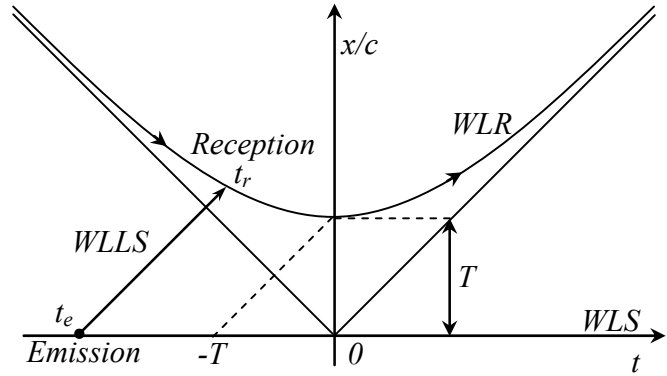

**Figure 2.** The world lines of the stationary source of light WLS, of the accelerating receiver WLR and of a light signal WLLS.

We present in Figure 2 the world line of an accelerating observer (WLR) described by (9), the world line of the stationary source (WLS) and the world line of a light signal emitted at a time $t_e$ (WLLS) described by

$$x = c \cdot (t - t_e) \qquad (12)$$

By intersecting the two world lines we obtain the reception time $t'_r$ measured by the moving observer as follows

$$cT \cosh \frac{t'_r}{T} = c\left(T \sinh \frac{t'_r}{T} - t_e\right) \qquad (13)$$

Solved for $t'_r$ (13) leads to

$$t'_r = -T \ln\left(\frac{-t_e}{T}\right) = T \ln\left(\frac{T}{-t_e}\right) \qquad -\infty < t_e < 0 \qquad (14)$$

Figure 2 and equation (14) tell us that the accelerating observer receives only the light signals that the source emits at negative times. The light signals emitted during the time interval $-\infty < t_e < -T$ are received at $t'_r < 0$ by the approaching observer who reduces his velocity. The light signals emitted in the time interval $-T < t_e < 0$ are received at $t'_r > 0$ by a receding observer who increases his velocity. As we see the relationship between $t'_r$ and $t_e$ has a non-linear character. If the stationary source emits periodic light signals at time instants $t_{e,N} = N \cdot T_e$ then the accelerating observer receives the $N^{th}$ light signal when his wrist watch reads $t'_{r,N} = -T \ln\left(\dfrac{-N \cdot T_e}{T}\right)$. Here $N$ is a negative integer number. Special relativity does not impose an upper limit to the magnitude of the acceleration. The emission frequency $\nu_e = 1/T_e$ can change in a very large range from mechanical frequencies, generated by periodically inserting and removing a shutter in the path of light emitted by the source, up to very high frequencies that approach the period of the electromagnetic oscillations in the wave.

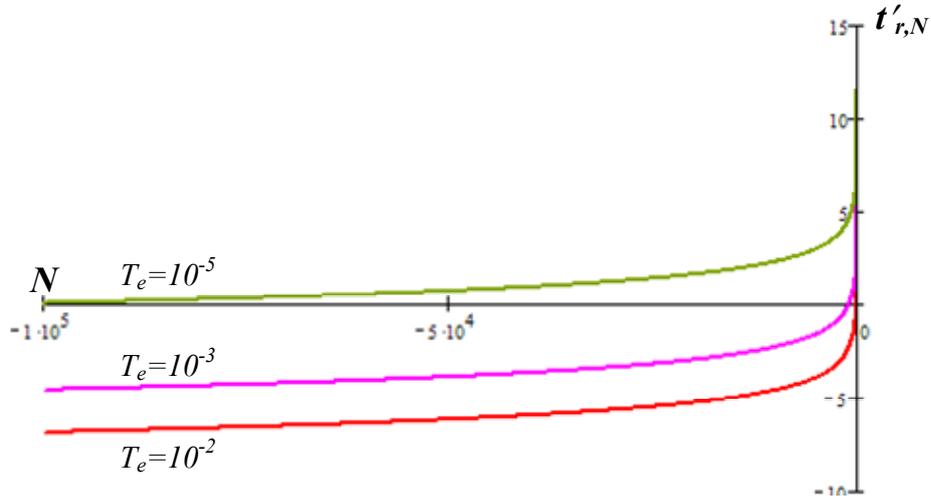

**Figure 3**. The variation of the time interval $t'_{r,N}$ between the reception of two successive light signals by the accelerating receiver vs. the order number $N$ of the light signal for $T=1$ and different values of $T_e$.

The time interval between the receptions of two successive light signals is given by
$$T'_{r,N} = T \ln \frac{N}{N-1} \qquad (15)$$
and we present its variation with $N$ in Figure 4. We obtain here a striking and very interesting result as the received period doesn't depend on the frequency (period) at which the stationary source emits the successive light signals. It depends however on the order number $N$.

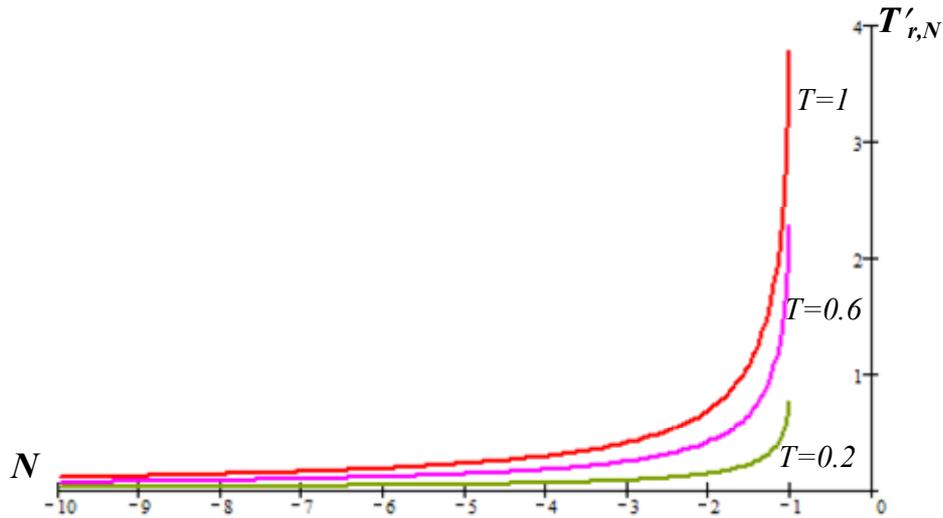

**Figure 4.** The variation of the time interval $T'_{r,N}$ between the reception of two successive light signals by the accelerating receiver vs. the order number $N$ of the light signal.

### 3. Stationary receiver and accelerating light source

We present this scenario in Figure 5. The receiver at rest at the origin $O$ receives the successive light signals emitted by the accelerating source.

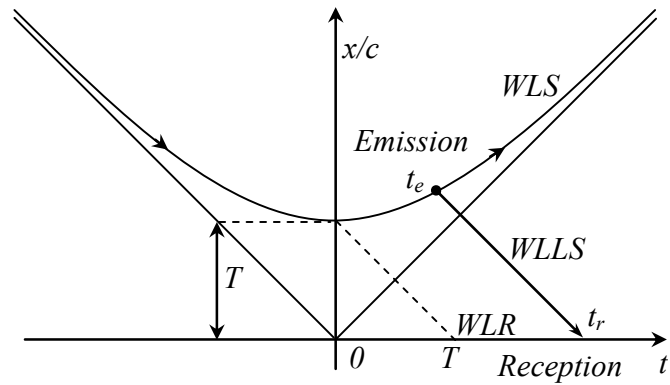

**Figure 5** The world lines of the accelerating light source WLS, of the stationary receiver WLR and of a light signal emitted by the source WLLS

As we see from Figure 5 the light signal emitted at a time $t'_e$ is received at a time $t_r$. In the time interval $0 < t_r < \infty$ the stationary observer receives all the light signals emitted by the accelerating source in the time interval $-\infty < N\tau'_e < \infty$. We obtain the relationship between them by intersecting the world line of the source with the world line of the light signal i.e.

$$cT \cosh\left(\frac{t'_e}{T}\right) = c\left(t_r - T \sinh\frac{t'_e}{T}\right). \qquad (16)$$

Solved for $t_r$ (16) leads to

$$t_r = T \exp\left(\frac{t'_e}{T}\right). \qquad (17)$$

Equation (17) reveals the non-linear character of the relationship between emission and reception times. With a signal source emitting periodic light signals at time instants $t'_{e,N} = N \cdot T'_e$ the receiver gets the $N^{th}$ signal at time $t_{r,N} = T \exp\left(\frac{NT'_e}{T}\right)$. We present in Figure 6 the variation of $t_{r,N}$ with $N$ for $T=1$ and different values of the emission frequency $v'_e = 1/T'_e$.

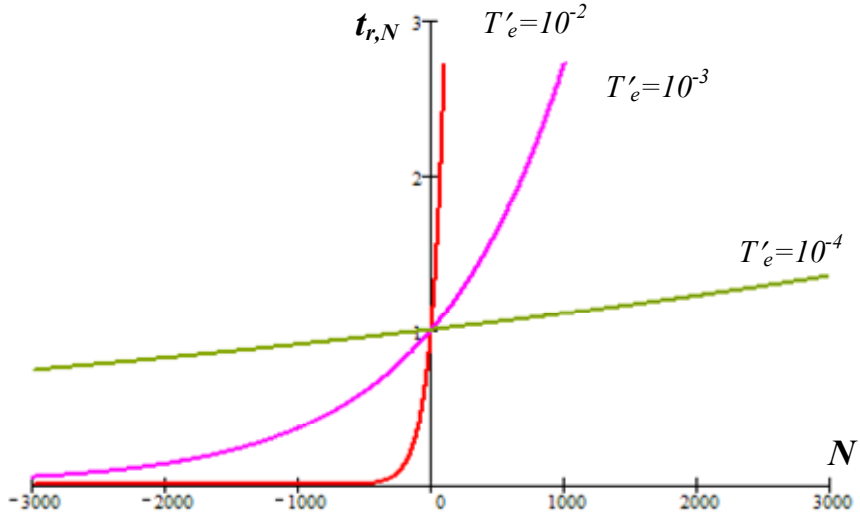

Figure 6. The variation of $t_{r,N}$ with $N$ for $T=1$ and different values of $T'_e$

The period at which the stationary observer receives the successive light signals is given by

$$T_{r,N} = T \exp\left(\frac{NT'_e}{T}\right)\left[1 - \exp\left(-\frac{T'_e}{T}\right)\right] \qquad (18)$$

and we present its variation with $N$ in Figure 7 for $T=1$ and for different values of the proper emission frequency $v'_e$.

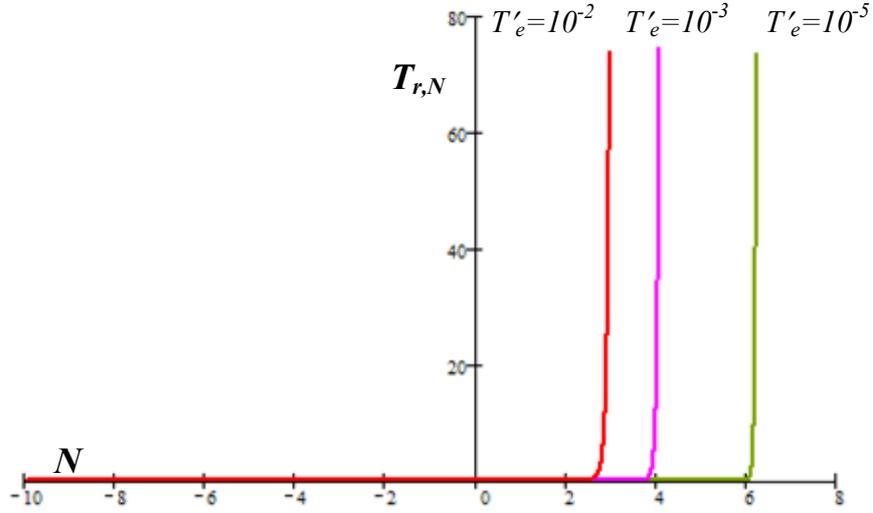

Figure 7. The variation of $T_{r,N}$ with $N$ for $T=1$ and different values of $T'_e$

Additionally (17) allows to calculate the emission time which corresponds to every particular reception time as: $t'_e = T \ln\left(\frac{t_r}{T}\right)$. Now suppose that the accelerating light source is triggered to emit short signals at particular time instances $t'_{e,N}$ such that the stationary receiver detects these signals at equidistant time intervals $t_{r,N} = N \cdot T_r$. In this case the emission period between two successive signals will be $T'_{e,N} = T \ln \frac{N}{N-1}$. Similar to (15) we observe the same striking effect namely that the emission period doesn't depend on the received period but depends on the order number $N$.

Summarising the results obtained above, we consider the symmetry of the two scenarios concerning the reciprocity of the acceleration motion as reflected by the relationship between emission and reception times. By reformatting (14) and (17) we found the following correspondence:

|  | Reception time | Emission time |
|---|---|---|
| Accelerated receiver | $\frac{-t'_r}{T} = \ln\left(\frac{-t_e}{T}\right)$ | $\frac{-t_e}{T} = \exp\left(\frac{-t'_r}{T}\right)$ |
| Accelerated source | $\frac{t_r}{T} = \exp\left(\frac{t'_e}{T}\right)$ | $\frac{t'_e}{T} = \ln\left(\frac{t_r}{T}\right)$ |

## 4. Stationary source of acoustic waves and accelerating receiver

The scenario we follow involves a stationary source of plane acoustic waves or a machine gun[9] that fires successive bullets at constant time intervals $T_e$ and an accelerating receiver who performs the motion described by (9). Figure 8 presents the world line of the accelerating receiver (WLR), the world line of the stationary machine gun (WLMG) and the world line of the bullet (WLB) emitted at a time $t_e$. R receives the signals for two times (events 1 and 2).

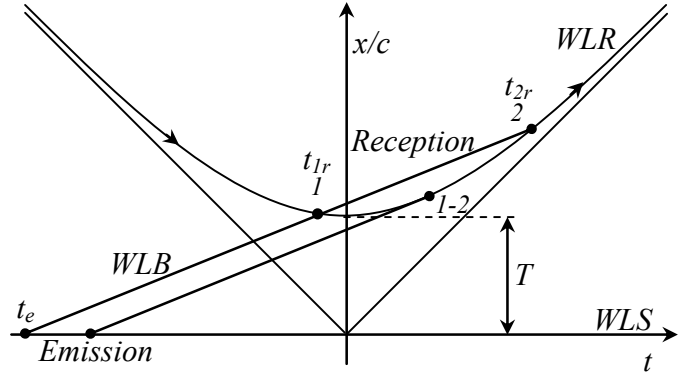

**Figure 8.** The world lines of the stationary machinegun WLMG, of the accelerating receiver WLR and of the bullet WLB.

In order to find out a relationship between the emission and reception times we intersect the world lines WLR and WLB. Similar to (13) we have

$$cT \cosh\left(\frac{t'_r}{T}\right) = u\left(T \sinh\frac{t'_r}{T} - t_e\right) \tag{19}$$

$u$ representing the propagation velocity of the acoustic wave or of the fired bullets. Solved for $t'_r$ and using the notation $\beta = u/c$ it leads to

$$t'_{1r} = T \ln\left[\frac{-\beta t_e + \sqrt{(\beta t_e)^2 - (1-\beta^2)T^2}}{(1-\beta)T}\right] \tag{20}$$

$$t'_{2r} = T \ln\left[\frac{-\beta t_e - \sqrt{(\beta t_e)^2 - (1-\beta^2)T^2}}{(1-\beta)T}\right] \tag{21}$$

As expected at limit when $u \to c$ we have $t'_{1r} \to T \ln\left(\frac{T}{-t_e}\right)$ as known from (14) and $t'_{2r} \to +\infty$.

When following the time evolution of the accelerating receiver R on figure 8 we distinguish the following characteristic situations depending on the values of *u/c*:
- WLB doesn't intersect WLR at all and thus no information change between source and receiver takes place,
- WLB is tangent to WLR (event **1-2)** characterized by the emission time

$$t_{1-2e} = -T\sqrt{\frac{1}{\beta^2} - 1} \qquad (22)$$

and the reception time when R receives the last light signal

$$t'_{1-2r} = T \ln\sqrt{\frac{1+\beta}{1-\beta}} \qquad (23)$$

As expected with $u \to c$ we have $t_{1-2e} \to 0$ and $t'_{1-2r} \to +\infty$.
- WLB intersects for two times WLR at the times $t'_{1r}$ and $t'_{2r}$ respectively. If the stationary source emits periodical signals at time instants $t_{e,N} = N \cdot T_e$ then the approaching observer R receives these signals (bullets or the wave crests) in the succession *–N,-(N-1)...-1, 0* at times $t'_{1r,N}$ after which he receives the same ones again in the reversed order at times $t'_{2r,N}$.

We present in Figure 9 the variation of the two reception times with *N* for *T=1* and different values of the emission frequency $T_e$.

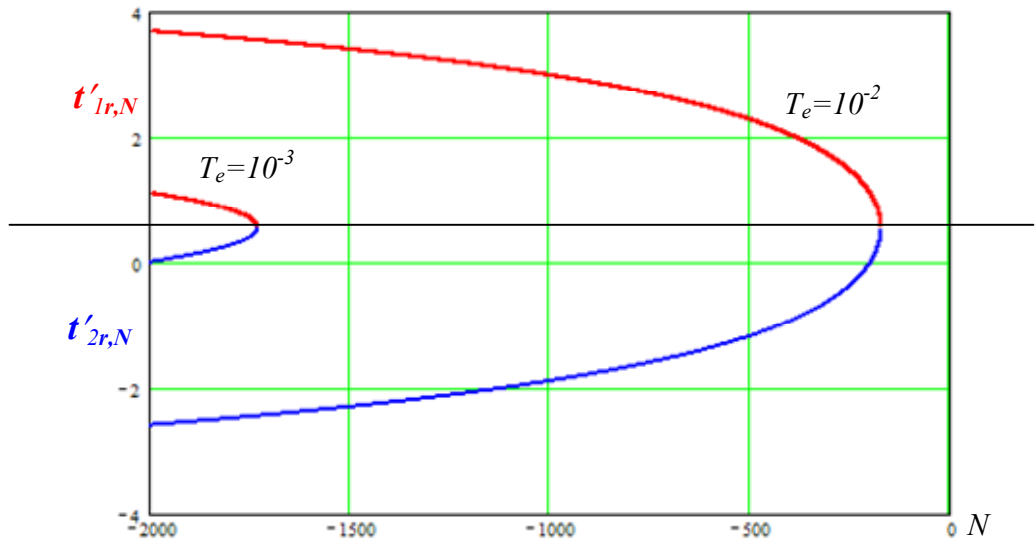

Figure 9. The variation of $t'_{1r,N}$ and of $t'_{2r,N}$ with *N* and the emission frequency $T_e$

## 5. Period (frequency) dependence

In order to be able to measure the period, the observer should receive at least two successive wave crests (light signals). In the case when the period is "very small" we can consider that the observer receives the two successive light signals (wave crests) being located at the same point in space. Under these conditions he is able to perform a continuous recording of the period. Returning to (7), which holds in the case of very small periods, we can recover Hamilton's[10] result by replacing $V$ with its instantaneous value given by (10) as it is shown by Cochran[3]. We could also start with (13) or with (19) allowing a continuous variation of the time coordinates. Differentiating both sides of the obtained equations we recover Cochran's results.

Of course all those results hold only in the case of the "very small" periods. Differentiating both sides of the equations (13) and (16) we recover Cochran's results.

Moreau[4] calls that situation "locality" in the period measurement. In the realistic case when the observer receives the two successive light signals being located at two different points in space he speaks about "non-locality" in the period measurement.

Non-locality is favoured by high velocities and large periods (low frequencies). In the case of an accelerating motion high velocities are achieved after long times of motion. As we have mentioned above, long emission periods can be achieved by periodically removing a shutter from the path of an electromagnetic wave for very short time intervals. The non-linearity of our equations (14) and (17) tell us that in the experiments described above different frequencies are differently shifted.

We could also characterize the effects studied above introducing the concept of Doppler factor D defined as the quotient between the proper reception period and the proper emission period. In the case of the stationary observer and accelerating source studied above (18) it is given by

$$D = \frac{T'_e}{T \exp\left(\dfrac{NT'_e}{T}\right)\left[1 - \exp\left(-\dfrac{T'_e}{T}\right)\right]}. \qquad (24)$$

## 6. Conclusions

Extending the approaches of Cochran[3] and of Moreau[4] to the problem of period measurement by accelerating observers, we reveal some unknown aspects of the problem, as the variation of the period with the order number of the emitted wave crest or the invariance of the emission and the reception periods in respect to each other. The case of the stationary source of acoustic waves and accelerating receiver is also studied revealing the change in the order in which the successive wave crests are received. The case of the stationary receiver and accelerating source of acoustic waves remains open due to the fact that the velocity of the emitted bullet depends on the velocity of the machine gun. Our approach underlines the importance of taking into account the "non-locality" in the period measurement by accelerating observers. Moreau[4] considers only the case of the period that separates the wave crests emitted for $N=1$ and $N=0$ respectively.